\documentclass{jnmp01b}

\usepackage{amsmath}

\numberwithin{equation}{section}

\allowdisplaybreaks

\def\dsp{\displaystyle}
\def\d{\mbox{\rm d}}
\def\e{\mbox{\rm e}}

\def\f{\frac}
\def\fei{\mbox{$\frac{5}{8}$}}
\def\fra#1#2{\mbox{$\frac{#1}{#2}$}}
\def\ha{\mbox{$\frac{1}{2}$}}
\def\oei{\mbox{$\frac{1}{8}$}}
\def\oqr{\mbox{$\frac{1}{4}$}}
\def\osi{\mbox{$\frac{1}{6}$}}
\def\oth{\mbox{$\frac{1}{3}$}}
\def\otw{\mbox{$\frac{1}{12}$}}
\def\p{\dsp\partial}
\def\tha{\mbox{$\frac{3}{2}$}}

\def\ga{\alpha}
\def\gd{\delta}
\def\gg{\gamma}
\def\gl{\lambda}
\def\gs{\sigma}
\def\gW{\Omega}

\def\bfa{{\bf a}}
\def\bfe{{\bf e}}

\def\pr{{\p\over\p r}}
\def\pt{{\p\over\p t}}
\def\pu{{\p\over\p u}}
\def\pv{{\p\over\p v}}
\def\pw{{\p\over\p w}}
\def\px{{\p\over\p x}}
\def\py{{\p\over\p y}}

\def\dddot#1{\mathinner{\buildrel\vbox{\kern5pt\hbox{...}}\over{#1}}}
\def\dm#1#2{\f{\dsp \d #1}{\dsp \d #2}}
\def\pa#1#2{\f{\dsp \p#1}{\dsp \p#2}}

\def\etal{{\it et al }}
\def\({\left (}
\def\){\right )}
\def\ie{{\it ie }}
\def\lb{\left[ }
\def\rb{\right] }
\def\re#1{(\ref{#1})}
\def\viz{{\it viz }}

\def\n{}
\def\z{&=}

\makeatletter
\DeclareRobustCommand{\primfrac}[1]{%
  \PackageWarning{amsmath}{%
Foreign command \@backslashchar#1; %
\protect\frac\space or \protect\genfrac\space should be used instead%
  }
  \global\@xp\let\csname#1\@xp\endcsname\csname @@#1\endcsname
  \csname#1\endcsname
}
\makeatother

\begin{document}

\renewcommand{\evenhead}{P.G.L.\ Leach, S Cotsakis and G.P.\ Flessas}
\renewcommand{\oddhead}{Symmetry, Singularities and Integrability in
Complex Dynamics I}


\thispagestyle{empty}

\begin{flushleft}
\footnotesize \sf
Journal of Nonlinear Mathematical Physics \qquad 2000, V.7, N~4,
\pageref{firstpage}--\pageref{lastpage}.
\hfill {\sc Article}
\end{flushleft}

\vspace{-5mm}

\copyrightnote{2000}{P.G.L.\ Leach, S Cotsakis and G.P.\ Flessas}

\Name{Symmetry, Singularities and Integrability in Complex Dynamics I:
The Reduction Problem}

\label{firstpage}

\Author{P.G.L.\ LEACH~$^{\dag\,\ddag}$, S COTSAKIS~$^\dag$ and G.P.\ FLESSAS~$^\dag$}

\Adress{$^\dag$ GEODYSYC, Department of Mathematics\\
~~University
of the Aegean, Karlovassi 83 200, Greece \\[2mm]
$^\ddag$ School of
Mathematical and Statistical Sciences, University of Natal,\\
~~Durban, South
Africa 4041}

\Date{Received February 3, 2000; Revised May 12, 2000;
Accepted May 13, 2000}

\begin{abstract}
\noindent
Quadratic systems generated using
Yang-Baxter equations are integrable in a sense, but we display a deterioration
in the possession of the Painlev\'e property as the number of equations in each
`integrable system' increases.  Certain intermediate systems are constructed
and also tested for the Painlev\'e property.  The Lie symmetries are also
computed for completeness.
\end{abstract}


\section{Introduction}

Golubchik and Sokolov \cite{Golubchik} recently discussed systems of ordinary
differential equations which are integrable by the standard factorisation
method of Adler-Kostant-Symes \cite{Kostant} or the generalised factorisation
method \cite{Golubchik2}.  They established relationships between such
reductions, operator Yang-Baxter equations and some kinds of nonassociative
algebras.  In their paper \cite{Golubchik} a number of specific examples was
given without much detailed comment on the solutions except for the
possibility of the system passing or not passing the Painlev\'e test.  The
examples are as follows:
\begin{equation}
\begin{split}
&P_t = P^2-RP-QS \\
&Q_t = (\beta-2)RQ + \beta PQ \\
&R_t = R^2 - RP -QS \\
&S_t = (3-\beta)RS + (1-\beta)PS
\end{split}
\label{1}
\end{equation}
\begin{equation}
\begin{split}
&P_t = RQ \\
&Q_t = PQ \\
&R_t = \ga RP
\end{split}
\label{2}
\end{equation}
\begin{equation}
\begin{split}
&P_t = 2PR + \gl QR \\
&Q_t = 2QR - \gl PR \\
&R_t = P^2 + Q^2 + R^2
\end{split}
\label{3}
\end{equation}
\begin{equation}
\begin{split}
&P_t = (\nu - \mu)QR \\
&Q_t = 2\mu PQ + \mu Q^2 + \nu QR \\
&R_t = -2\nu PR - \nu R^2 - \mu QR
\end{split}
\label{4}
\end{equation}
and
\begin{equation}
\begin{split}
&\dm{X}{t} = X^2 + 3XY = X(X+3Y) \\
&\dm{Y}{t} = 3XY + Y^2 = Y(3X+Y)
\end{split}
\label{5}
\end{equation}
in which Greek letters are constants.  We have given the sets of equations in
the order and the form as presented by \cite{Golubchik}.

The integrability of these systems, apart from the two-dimensional system in
\re{5}, is not immediately obvious, although in the paper of Golubchik and
Sokolov \cite{Golubchik} we are informed that they are so constructed to be
integrable in the sense of the generalised factorisation method.  Unfortunately
they do not connect this definition of integrability with the
usual ones which are as follows:
\begin{description}
\item
(a) The existence of an explicit solution relating the variables, not necessarily
the dependent variables as explicit functions of the independent variable, or,
in the instance of systems, a number of explicit independent functions equal to the
number of the dependent variables;
\item
(b) The existence of a sufficient number of independent first integrals and
invariants which could be used to give a local version via the implicit
function theorem of (a) above;
\item
(c) The existence of a sufficient number of symmetries either to reduce the
differential equations of the system to algebraic equations or to obtain the
independent first integrals and invariants of (b) above;
\item
(d) The passing of the Painlev\'e test, either in the normal form or the weak form,
which, in the form as given by Conte \cite{Conte}, will guarantee the
existence of a solution as required by (a) above that is analytic in the
complex $ t $-plane or in a region of it apart from isolated moveable
singularities which are poles or algebraic branch points.
\end{description}
The only hint which the two authors give is their comment associated with
\re{1} to the effect that for general values of the parameter $ \beta $, one
would not expect the system to pass the Painlev\'e-Kovalevskaya test.  One
infers from this that they are not looking for a solution which is analytic
apart from isolated singularities which are either poles or algebraic
and so one can presume that they intend (a) above.

There are those who would insist that a solution be a (single-valued) function
(cf Conte \cite{Conte}) and, in the strict
mathematical sense, they are quite correct.  However, in practice the domain of
the variable, $ t $, in which the solution is going to be used is much smaller
than the complex $ t $-plane.  It is with this emphasis on the practicality of
applications that (a) above is proposed.

It is the intention of this paper to take up the systems enumerated above
where the authors of \cite{Golubchik} left them.  We shall see that the
solutions of these systems of equations have certain interesting aspects
which illustrates various properties of differential equations.

The most
obvious one is that all of the systems quoted are quadratic systems which
are of tremendous interest in many areas of mathematical modelling ranging
from studies of population systems to the reactions in chemical systems.
They, along with the closely related Lotka-Volterra systems, are usually
studied from the point of view of dynamical systems since in general they
are not integrable, but are well adapted for intensive and informative
qualitative investigation.  Nevertheless there has been a number of studies
of these systems and the circumstances under which they possess first integrals
and/or invariants \cite{Cairo1,Cairo2,Cairo3}.  The connection between the
possession of the Painlev\'e property and the existence of first
integrals/invariants has also been investigated \cite{Hua}.

A less obvious property of the systems
presented by Golubchik and Sokolov is that
all of them possess the two point symmetries
of invariance under time translation and the selfsimilar transformation.
These two symmetries immediately make them interesting from the point of
view of the Painlev\'e property since there is a very close relationship
between the nature of the singularity and the specific form of the
selfsimilar symmetry \cite{Feix}.  The selfsimilar symmetry permits an
analysis of the next to leading order behaviour and the Painlev\'e property
follows from this when the next to leading order behaviour becomes a Laurent
expansion about the singularity \cite{Feix2}.

There is the question whether the possession of the two symmetries mentioned
above is
sufficient to guarantee integrability.  As we have been informed in
\cite{Golubchik}, all of
these systems are integrable, we have an excellent opportunity to examine the
symmetry structure of these systems and equations related to them in the
knowledge that the results will characterise integrable systems.  Of course
it is an open question whether the examination will lead to a clear result.
We defer our comments on the question until the conclusion.  However, our
path to the conclusion will reveal a number of interesting features of these
systems and their connections to other well known facets of integrable
systems.

On the presumption that complexity of appearance is a representative of
complexity of fact we shall more or less work backwards through the list of
systems above and attempt to examine them in increasing order of difficulty.
In the case of some of the systems we make use of rescaling to make them look
simpler.  This may simply be a psychological trick, but it is one worth
practising on all occasions. One of our principal activities will be the
examination of these systems and any derivate equations for their Lie
symmetries.  As an aid in this examination we make use of Program
Lie \cite{head, head2} to attend to the dirty business of the calculation of the
symmetries.  This program, which is one of the more successful analysers for
symmetry, is freely available on Simtel sites.  Naturally we shall also be
concerned with the possession of the Painlev\'e-Kovalevskaya property either
by a direct computation or by a consideration of the structure of the
solution of the system of equations.

Before we commence our analysis we mention some terminology introduced in
reference \cite{Leach2}.  In systems of differential equations invariant under
time translation and rescaling all terms contribute to the dominant behaviour.
However, it is possible that a subset of terms has an additional or different
rescaling property.  This symmetry is termed `subselfsimilarity' and the terms
possessing this symmetry are called `subdominant'.  These terms also have to be
analysed for possession of the Painlev\'e property, naturally with compability
at the resonances no longer guaranteed.  A simple example of the possession of
subselfsimilarity is the generalised Chazy equation
\begin{equation}
\dddot{x} + ax\ddot{x}+b\dot{x}^2 = 0.  \label{5a}
\end{equation}
The equation as a whole has the two symmetries $G_1 = \p/\p t$ and $G_2 =
-t\p/\p t + x\p/\p x$.  The second and third terms have the two homogeneity
symmetries $t\p/\p t$ and $x\p/\p x$ and so would also have to be analysed
separately.

Finally we note that this is the first in a series of papers devoted to the
intertwined subjects of symmetry, singularity and integrability.  The second
paper in the series \cite{paper2} treats the singularity analysis and
integrability of the class of
second order equations invariant under time-translation and rescaling and the
third \cite{paper3} some aspects of complete symmetry groups \cite{Krause}.

\section{The two-dimensional system}

\subsection{Painlev\'e analysis}

We rewrite  \re{5}, \viz
\begin{equation}
\begin{split}
&\dm{X}{t} = X^2 + 3XY \\
&\dm{Y}{t} = 3XY + Y^2
\end{split}
\label{2.1}
\end{equation}
as
\begin{subequations}
\label{2.2}
\begin{gather}
\dot{x} = x^2 + 3xy \\
\dot{y} = 3xy + y^2.
\end{gather}
\end{subequations}

When we perform
the Painlev\'e analysis on the system \re{2.2},
 we find that, using
the usual notations and methods, the singularity
corresponding to the selfsimilar symmetry of the system gives the following
behaviour
\begin{equation}
\begin{split}
&x = -\oqr\tau^{-1} + 3\mu\tau^{-\ha},\quad p = -1,\quad r=-1,\ha \\
&y = -\oqr\tau^{-1} - 2\mu\tau^{-\ha}, \quad q = -1,\quad r=-1,\ha,
\end{split}
\label{2.11}
\end{equation}
where $ \tau=t-t_0 $ and $ t_0 $ is the location of the movable singularity
and $ \mu $ is the second arbitrary constant of integration, which indicates the
weak Painlev\'e property implied in the solution (see \re{2.10}).

There is also the
possibility of a different type of singular behaviour for \re{2.2}
which would be
given by the two equivalent possibilities $ p=-1,q>-1 $ and $ p>-1,q=-1 $.
This type of behaviour is found for example in the analysis of the Mixmaster
universe \cite{Contopoulos,Contopoulos2}.  A simple-minded approach would be to write
\begin{equation}
x = \sum_{i=0}^{\infty}a_i\tau^{i-1},\quad y = \sum_{i=0}^{\infty}\tau^i
\label{c1}
\end{equation}
for the first of these possibilities.  This yields the particular solution
\begin{equation}
x = \frac{-1}{\tau}, \qquad y = 0. \label{c2}
\end{equation}
However, bearing in mind the expansion in powers of $\tau^{\ha}$ found in
\re{2.11} we do well to propose
\begin{equation}
x = \sum_{i=0}^{\infty}a_i\tau^{\ha i-1},\quad y = \sum_{i=0}^{\infty} \tau^{\ha
i-\ha}.  \label{c4}
\end{equation}
This gives the one-parameter solution
\begin{equation}
\begin{split}
&x = -\osi\tau^{-1}+\osi\tau^{-\ha}-\mbox{$\frac{13}{54}$}+\ldots \\
&y = -\oth\tau^{-\ha}-\mbox{$\frac{1}{9}$}+
\mbox{$\frac{7}{27}$}\tau^{\ha}+\ldots.
\end{split}
\label{c5}
\end{equation}
Since \re{c5} is a one-parameter system, this pattern of singular behaviour
(and clearly the one corresponding to the interchange of the roles of $x$ and
$y$) does not pass the weak Painlev\'e test and so the system \re{2.2} does not
possess the weak Painlev\'e Property.  (In this we follow the definition of
Tabor \cite{Someone}[p 330].)

\subsection{First reduction}

We now obtain the solution of \re{2.2} as follows.  Firstly we
are immediately struck by the resemblance to the coefficients in the
expansion of a cubic
binomial.  This suggests the change of variables
\begin{equation}
u = x^2+y^2\qquad v=x+y \label{2.3}
\end{equation}
so that the system  \re{2.2} takes the form
\begin{subequations}
\label{2.4}
\begin{gather}
\dot{u} = 2v^3 \\
\dot{v} = -2u+3v^2
\end{gather}
\end{subequations}
which by differentiation of the second of these and substitution for
 $ \dot{u} $ gives the second order equation
\begin{equation}
\ddot{v}-6v\dot{v}+4v^3 = 0. \label{2.5}
\end{equation}
Equation  \re{2.5} is of a type \cite[6.43, p 551]{Kamke}
which has been studied extensively in the
literature because of the frequency of its occurrence in different
areas \cite{Mohammed1,Mohammed2,Mohammed3,Mohammed5,Mohammed6}
and in particular is known, for certain values of the coefficients, to have
eight symmetries instead of the obvious two of invariance under time
translation and selfsimilarity, thereby making it
linearisable \cite{Mohammed4}.

For the system \re{2.4} we obtain two possible sets of leading order behaviour.
They are $p=-2$ and $q=-1$ with the resonances $r=-1,1$ and $p=q=-1$ with the
resonances $r=-1,-2$.  We give the critical
parts of the expansions.  They are
\begin{equation}
\begin{split}
&u = \oei\tau^{-2} + 3\mu\tau^{-1}+\ldots \\
&v = -\ha\tau^{-1} -2\mu+\ldots,
\end{split}
\label{2.12}
\end{equation}
which gives the expansion about the singularity at $ t=t_0 $, and
\begin{equation}
\begin{split}
&u = t^{-1} + 2\mu t^{-2} + 3\nu t^{-3} \\
&v = t^{-1}+\mu t^{-2}+\nu t^{-3},
\end{split}
\label{2.13}
\end{equation}
which represents the asymptotic expansion of the solution.  The second constant
of integration is $ \nu $ and we have followed the practice introduced by Feix
\etal \cite{Feix} of making the expansion in the variable $ t $ because of the
asymptotic nature of the expansion.

The expansion in  \re{2.12} has been termed \cite{Feix} a right Painlev\'e series
because it is a Laurent expansion about the singularity at $ t_0 $ in ascending
powers of $ \tau $ whereas that in  \re{2.13} is called a left Painlev\'e series
because it is a Laurent expansion in descending powers of the independent
variable, $ t $.  The existence of what is now called the left Painlev\'e series,
that is an asymptotic series as a representation of the solution, has only
been recognised in recent years \cite{Lemmer}.

It is interesting that the
subselfsimilarity symmetry property of the system  \re{2.2} and its
corresponding Painlev\'e analysis disappears under the transformation which
leads to the system  \re{2.4}.  The reason for this is obvious for the first
of  \re{2.4} contains only two terms and a minimum of three terms is required
for the presence of a subselfsimilarity symmetry \cite{Feix} (these can be
additive or multiplicative).  It is an
interesting speculation as to whether there is always a transformation which
will remove the possibility of the subselfsimilar behaviour.

\subsection{Second reduction}

An alternative route to the analysis of \re{2.2} initiated by the
transformation \re{2.3} is given by the transformation
\begin{equation}
u = x + y,\quad v = x - y. \label{2.14a}
\end{equation}
The transformed system \re{2.2} is now
\begin{subequations}
\label{2.15a}
\begin{gather}
\dot{u} = 2u^2 - v^2 \\
\dot{v} = uv.
\end{gather}
\end{subequations}
We have two possible routes for the solution of \re{2.15a}, \viz the
differentiation of (\ref{2.15a}a) and elimination of $v$ or the differentiation
of (\ref{2.15a}b) and elimination
of $u$.
The first route produces the nonlinear second order ordinary differential
equation
\begin{equation}
\ddot{u} - 2u\dot{u} - 4u^3 = 0 \label{2.16a}
\end{equation}
which belongs to the same class of equations as \re{2.5}. If we perform the
leading order analysis on \re{2.16a}, we find that the only possible singularity
is a simple pole since there is no possibility of subselfsimilarity.
The coefficient of the leading power can be either $-\ha$ or $1$. In the
former case the
resonances are at $r = -1, 3$ and in the latter case at $r = -1, 6$.
There is no
possibility of inconsistency at the second resonance because of the rescaling
symmetry. Both expansions are right Painlev\'e series and the equation
possesses the Painlev\'e property. If we look to the solution of \re{2.16a}
by means
of reduction of order, we come to an Abel's equation of the second kind and
firstly transformation is necessary to obtain the solution. An alternative route
is to use the Riccati transformation of \re{2.6} (with the same value of $\ga$)
to obtain the third order equation
\begin{equation}
w\dddot{w}-2\dot{w}\ddot{w}=0 \label{2.17a}
\end{equation}
which has a solution in terms of elliptic functions.

The second route produces the nonlinear second order ordinary differential
equation
\begin{equation}
v\ddot{v} - 3\dot{v}^2 + v^4 = 0 \label{2.18a}
\end{equation}
which belongs to a class of equations to be found in Kamke \cite[6,128, p
574]{Kamke}.
Under the transformation $v = w^{-\ha}$ \re{2.18a} takes the particularly
simple form
\begin{equation}
\ddot{w} = 2 \label{2.19a}
\end{equation}
which we shall again encounter in \re{3.3}. The solution of \re{2.19a} is
trivial and so we have
\begin{equation}
v(t) = (A + 2Bt + t^2)^{-\ha}. \label{2.20a}
\end{equation}
The other solution $u(t)$ follows easily from (\ref{2.15a}b). One
easily checks that these two solutions lead to the same forms for
the general solutions of the original system, \re{2.2}.

Equation \re{2.16a} has just the two Lie point symmetries of invariance under
time translation and selfsimilarity. In marked contrast \re{2.18a} possesses
eight
Lie point symmetries which, in a sense, makes it more like \re{2.5} than
\re{2.16a} is. We observe that the first two terms of \re{2.18a} are invariant
under the
homogeneity symmetry $u\p/\p u$ and so expect a more complicated pattern of
leading order behaviour. This is borne out by the analysis which gives in the
case of all terms dominant an exponent of $-1$ with coefficients $\pm 1$
and in the
subdominant case exponents of 0 and $-\ha$. In the former case the resonances
are given by $r = -1, -2$ and so for this type of leading order behaviour the
equation passes the Painlev\'e test and has a left Painlev\'e series,
that is, an
asymptotic solution. This is easily seen from \re{2.20a} if we expand it in the
asymptotic form
\begin{equation}
v(t) = \f{1}{t}\(1-\f{B}{2t}+\f{3B^2-4A}{8t^2}+\ldots\). \label{2.21a}
\end{equation}
For the subdominant behaviour and $p = 0$ the first few terms of the Taylor
expansion give
\begin{equation}
v(t) = a_0 + a_1t + \f{1}{2a_0}\(3a_1^2 - a_0^4\) t^2 + \(1 - 3a_0^2\) a_1t^3
+ \ldots. \label{2.22a}
\end{equation}
We observe that the third term in \re{2.18a}, which breaks the homogeneity
symmetry of the first two terms, does not affect the existence of the Taylor
series solution. The coefficients of the first two terms are arbitrary constants
and so this is the series representation of a general solution. In the case that
$p = -\ha$ we obtain
\begin{equation}
v(t) = a_0\tau^{-\ha} - \ha a_0^3\tau^{\tha} - \fei a_0^4\tau^{\fra{7}{2}}2 + \ldots.
\label{2.23a}
\end{equation}
This series representation of the solution contains the two arbitrary constants
$a_0$ and $t_0$. The equation satisfies the requirements of the weak Painlev\'e
test. This is one occasion when all possible leading order behaviours
possess
the Painlev\'e property and so there is no question that the equation \re{2.18a}
possesses the Painlev\'e property.
For the sake of completeness, not to mention a little amusement, we
perform the Painlev\'e analysis of the system \re{2.15a}. With the usual
substitution for the leading order behaviour the exponents are
\begin{equation}
\begin{array}{rll}
p-1& 2p& 2q\\
q-1& p+q&
\end{array} \label{2.24a}
\end{equation}
so that we have the possibility of all terms being dominant with $p = q = -1$
or of just one variable being dominant with $p = -1, q > -1$. In the former
case the resonances occur at $r = -1, -2$ and so we have a left Painlev\'e
series.  Because of the symmetry of the system we shall not have
inconsistencies at
the resonances and we obtain the first few terms of the expansion as
\begin{equation}
\(\begin{array}{cc}
u\\v \end{array}\) = \(\begin{array}{rr}
-1\\ \beta \end{array}\)t^{-1}+\(\begin{array}{rr}
1\\-\beta \end{array}\)\mu t^{-2}+\(\begin{array}{rr}
-2\beta\\ 1 \end{array}\)\nu t^{-3} +\ldots, \label{2.25a}
\end{equation}
where $\mu$ and $\nu$ are the two arbitrary constants
required for a general solution.
For the second case the substitution of the series expansion
\begin{equation}
u = \sum_{i=0}^{\infty}a_i\tau^{i-1}\quad v = \sum_{i=0}^{\infty}
b_i\tau^i \label{2.26a}
\end{equation}
results in the particular solution
\begin{equation}
u(t) = -\f{1}{2\tau}\quad v(t) = 0. \label{2.27a}
\end{equation}
We thus arrive at the interesting conclusion that, while we started with a
system, \re{2.2}, which does not possess the (weak) Painlev\'e property, we can map it
through the nonlinear transformation \re{2.3} to the new system \re{2.4}
that has the strong Painlev\'e property and through the linear nonsingular
transformation \re{2.14a} to the new system \re{2.15a} which, although it
has the (strong) Painlev\'e property, naturally splits into
two single equations, \viz \re{2.16a} and \re{2.18a}, one possessing the (strong)
Painlev\'e property and the other the weak Painlev\'e property.

\subsection{General solution}

It so happens that \re{2.5} has the combination of coefficients which gives
eight Lie point symmetries.  In principle one can use these symmetries to
obtain the solution, but they are particularly complicated and there is an
easier route to the solution.  It is known \cite{Mohammed1} that the equations
which belong to this class can be expressed as a simple third order equation
by means of a Riccati transformation.  We put
\begin{equation}
v = \ga\f{\dot{w}}{w}, \label{2.6}
\end{equation}
where $ \ga $ is a constant to be determined to give the third order equation
an optimal simplicity \cite{bkp?}.
Using  \re{2.6} we find that  \re{2.5} becomes
\begin{equation}
\begin{split}
&\f{\dddot{w}}{w}-\(3+6\ga\)\f{\dot{w}\ddot{w}}{w^2}+\(2+6\ga+4\ga^2\)
\f{\dot{w}^3}{w^3}=0 \\
&\Leftrightarrow \dddot{w} = 0
\label{2.7}
\end{split}
\end{equation}
when we put $ \ga=-\ha $  for both of the terms containing $ \ga $  then vanish.
Equation  \re{2.7} is trivial to solve and we obtain
\begin{equation}
v(t) = -\f{B+Ct}{A+2Bt+Ct^2},\label{2.8}
\end{equation}
where $ A $, $ B $ and $ C $ are constants of integration.  Equation (\ref{2.4}a)
is now a simple quadrature and we obtain
\begin{equation}
u(t) =\f{2(B+Ct)^2-(B^2-AC)}{(A+2Bt+Ct^2)^2},\label{2.9}
\end{equation}
for which an additive constant of integration must be set at zero to maintain
consistency with (\ref{2.4}b).  (Despite the presence of the three constants
 $ A $, $ B $ and $ C $  they provide only two independent constants of
integration.)
Finally we obtain the solutions of the original system of equations.  We have
\begin{equation}
x(t) = -\ha\f{B+Ct \pm\sqrt{3(B+Ct)^2-2(B^2-AC)}}{A+2Bt+Ct^2} \label{2.10}
\end{equation}
and $ y(t) $ has the same form with the opposite sign attached to the square root.
We observe that in general the solutions $ x(t) $  and $ y(t) $  have branch
point singularities whereas the only singularities of $ u(t) $  and $ v(t) $
are always poles.

\subsection{Symmetry analysis}

We complete our considerations of the system  \re{2.2} by looking at its first
integral and invariant and some of its Lie point symmetries.  Since the
systems  \re{2.2} and  \re{2.4} (or \re{2.2} and \re{2.15a})
are related by a point transformation, the
discussion of one system is equivalent to the discussion of the other system
when it comes to Lie point symmetries and, as these underlie the existence of
the first integral and invariant, of them as well.  We have noted that all of
the systems discussed in this paper have the two Lie point symmetries which
leave them invariant under the transformations of time translation and
rescaling, that is
\begin{equation}
G_1 = \pt\quad\mbox{\rm and}\quad G_2 =- t\pt+x_i\pa{}{x_i},  \label{2.14}
\end{equation}
where the summation over $ i $  includes all of the dependent variables.  The
first integral is associated with $ G_1 $   and is simply obtained from the
integration of the ratio of (\ref{2.2}a) and (\ref{2.2}b), \viz
\begin{equation}
\dm{x}{y} = \f{x^2+3xy}{3xy+y^2}.  \label{2.15}
\end{equation}
It is
\begin{equation}
I = \f{(x-y)^4}{xy}. \label{2.16}
\end{equation}
We use $ G_2 $  to obtain the invariant from the autonomous first order ordinary
differential equation satisfied by the two characteristics, $ x=ut $
and $ y=vt $,
which is
\begin{equation}
\dm{v}{u} = \f{3uv+v^2+v}{u^2+3uv+u}.  \label{2.17}
\end{equation}
The integration of  \re{2.17} is not as simple as that of  \re{2.15} since for
the latter the integration is simply a quadrature whereas for the former a
first order ordinary differential equation has to be solved.  (This is
more or less the same as the situation as the integration of a second order
ordinary differential equation with the two symmetries  \re{2.14} \cite{Lie}.)
In terms of the original variables the invariant is
\begin{equation}
J = \f{t(x-y)^2+(x+y)}{(xy)^{\ha}} \label{2.18}
\end{equation}
in which we see that the coefficient of the independent variable, $ t $, is
simply $ I^{\ha} $.  In principle we could solve  \re{2.16} and  \re{2.18} for
 $ x $  and $ y $  as functions of $ t $.

The Lie point symmetries of the system  \re{2.2} are the equivalent of
generalised symmetries at the second order level and consequently there will
be an infinity of them.  We have no intention of presenting them all here!
We do note, however, that, if $ G $ is a Lie point symmetry of  \re{2.2}, then
so also is $ f(I,J)G $.  This is the equivalent result to that found in the
case of second order ordinary differential equations \cite{Leach}.  We can
limit the number of symmetries found by simply making an ansatz for the
functional dependence of the coefficient functions.  Thus, if we assume at
most quadratic dependence on the variables, we obtain the three symmetries
\begin{equation}
\begin{split}
G_1 \z \pt\n\\
G_2 \z -t\pt+x\px+y\py\n\\
G_3 \z \(x^2+3xy\)\px+\(3xy+y^2\)\py \label{2.19}
\end{split}
\end{equation}
and, if we allow cubic dependence, the further four symmetries
\begin{equation}
\begin{split}
G_4 \z t\lb \pt+\(x^2+3xy\)\px+\(3xy+y^2\)\py \rb \n\\
G_5 \z x\lb \pt+\(x^2+3xy\)\px+\(3xy+y^2\)\py \rb \n\\
G_6 \z y\lb \pt+\(x^2+3xy\)\px+\(3xy+y^2\)\py \rb \n\\
G_7 \z (x-y)^2\lb(3x+y)\px+(x+3y)\py\rb \label{2.20}
\end{split}
\end{equation}
are added.  We note that the power of the independent variable, $ t $, lags
behind the powers of the dependent variables.

\section{The simplest three-dimensional system}

\subsection{Painlev\'e analysis}

From a casual observation of the three three-dimensional systems
listed in the introduction one cannot, {\it a priori}, judge which
of them is the simplest. We shall presume that it is  \re{2} on
the basis that the system contains the fewest number of terms of
the three.  Our first step is to write the system in the simplest
form by rescaling the variables $ P $  and $ Q $  so that $ \ga $
becomes unity.  Thus we examine the system
\begin{subequations}
\label{3.1}
\begin{gather}
\dot{u} = vw \n\\
\dot{v} = wu \n\\
\dot{w} = uv
\end{gather}
\end{subequations}
which is a special case of the Rikitake System.

Because of the structure of the system  \re{3.1} there can be no subselfsimilar
symmetry and the leading order behaviour is that of a simple pole.  To make
up for this simplicity there are four possible sets of values for the
coefficients of the leading terms.  If we take the leading terms to be
 $ \ga\tau^{-1} $, $ \beta\tau^{-1} $  and $ \gg\tau^{-1} $, the possible
combinations are given in Table \ref{tab1}.
\begin{table}
\caption{\label{tab1}The four sets of values for the
coefficients of
the leading terms of the Right Painlev\'e Series for \re{3.1}.}
\centering
$
\begin{array}{r|r|r} \hline
  \ga  &  \beta  & \gg \\ \hline
-1 & 1 & 1\\
1 & 1 &-1\\
1 &-1 & 1\\
-1 &-1 &-1 \\ \hline
\end{array}
$
\end{table}
The resonances are determined from the solution of the characteristic
equation
\begin{equation}
\left|
\begin{array}{rrr}
r-1 &-\gg & -\beta\\
-\gg &r -1 &-\ga\\
-\beta &-\ga &r-1
\end{array}
\right|=0. \label{3.6a}
\end{equation}
They are $ r =-1, 2 (2) $.  The repetition of the resonance at $ r = 2 $  is not
a cause for concern since the matrix for the determination of the resonances
is a real symmetric matrix and so the geometric multiplicity of the eigen
vectors equals the algebraic multiplicity of the eigenvalues.  The expansion
up to the resonance is
\begin{equation}
\lb\begin{array}{c}
u\\v\\w
\end{array}\rb =\lb\begin{array}{c}
\ga\\ \beta\\ \gg
\end{array}\rb\tau^{-1}+\left\{\lb\begin{array}{c}
\gg \\ 1 \\ 0
\end{array}\rb\mu +\lb\begin{array}{c}
\beta \\ 0 \\ 1
\end{array}\rb\nu\right\}\tau\label{3.7}
\end{equation}
and so the system  \re{3.1} possesses the (strong) Painlev\'e property.

\subsection{First reduction and symmetry analysis}

Differentiation of (\ref{3.1}a) and the substitution of (\ref{3.1}b) and
(\ref{3.1}c), division by $ u $  and the same process of differentiation and
substitution gives the third order equation
\begin{equation}
u\dddot{u}-\dot{u}\ddot{u}-4\dot{u}u^3=0 \label{3.2}
\end{equation}
which, at the level of contact symmetries, only has the two symmetries
relating to time translation and rescaling.  The Painlev\'e analysis of
\re{3.2} reveals three patterns of leading order behaviour.  The all terms dominant
case has a simple pole and the resonances are $r= -1,2$ and $4$.  There is no
inconsistency at the second resonance and this case passes the Painlev\'e test.
Subselfsimilar behaviour can be found with the first two terms which share
the two homogeneity symmetries, $t\p/\p t$ and $u\p/\p u$.  The leading order
terms are in $\tau^0$ and $\tau^1$.  For both we obtain a Taylor series
containing three arbitrary constants.  Consequently \re{3.2} possesses the
(strong) Painlev\'e property.

We reduce the order of  \re{3.2}
using $ u^2 $  and $ \dot{u}^2 $, the two invariants of $ G_1=\p/\p t $, as the
new independent ($ x $) and dependent ($ y $) variables to obtain the second
order ordinary differential equation
\begin{equation}
y''=2. \label{3.3}
\end{equation}
Equation  \re{3.3} is a linear second order equation of very simple appearance.
This indicates that \re{3.2} has nonlocal symmetries in sufficient supply to
give the necessary right point symmetries for the second order equation.
Equation \re{3.3} has the solution
\begin{equation}
y(x)=A+Bx+x^2 \label{3.4}
\end{equation}
and so the solution of  \re{3.2} is
\begin{equation}
\pm(t-t_0)=\int\f{\d u}{\sqrt{A+Bu^2+u^4}} \label{3.5}
\end{equation}
in which the integral can be evaluated in terms of elliptic integrals, the
expression of which is not informative and so we omit it.  Because of the
cyclic symmetry of  \re{3.1} $ v $  and $ w $  are given by expressions of the
same form as  \re{3.5}.

It is a trivial matter to obtain the two autonomous integrals of the system
\re{3.1}.  They are
\begin{equation}
I = u^2-v^2\quad\mbox{\rm and}\quad J=u^2-w^2.  \label{3.6}
\end{equation}
The third cyclic expression is not independent of $ I $  and $ J $.  The
determination of the necessary invariant to complete the set (with $ I $
and $ J $) using the selfsimilar symmetry, $ G_2 $, seems to be impossible.
The one invariant which comes moderately easy is simply the ratio of $ I $
and $ J $  which is of no use.

As in the case of the two-dimensional system the system  \re{3.1} possesses an
infinite number of Lie point symmetries and we confine our attention to those
symmetries which are up to cubic in the variables for the coefficient functions.
Thus we have
\begin{gather}
G_1 = \pt\notag\\
G_2 = -t\pt+u\pu+v\pv+w\pw\notag\\
G_3 = vw\pu+wu\pv+uv\pw\notag\\
G_4 = t\lb \pt+vw\pu+wu\pv+uv\pw\rb\notag\\
G_5 = u\lb \pt+vw\pu+wu\pv+uv\pw\rb\notag\\
G_6 = v\lb \pt+vw\pu+wu\pv+uv\pw\rb \label{3.8}\\
G_7 = w\lb \pt+vw\pu+wu\pv+uv\pw\rb\notag\\
G_8 = \(u^2-v^2\)\pt\notag\\
G_9 = \(u^2-w^2\)\pt\notag\\
G_{10} = \(u^2-v^2\)\lb-t\pt+u\pu+v\pv+w\pw\rb\notag\\
G_{11} = \(u^2-w^2\)\lb-t\pt+u\pu+v\pv+w\pw\rb. \notag
\end{gather}
We observe that the two first integrals, $ I $  and $ J $, occur
as coefficients in $ G_8 $  to $ G_{11} $.  Although we noted the
possibility of this occurrence for the system  \re{2.2}, the form
of the first integral and invariant for that system was such that
symmetries of that type would not be found using the ansatz we
have adopted.   In the case of the system  \re{3.1} the two first
integrals fitted in with the ansatz and so we see the appearance
of symmetries containing them as common multipliers.  Given the
fact that we cannot determine the invariant for this system, the
likelihood of us making an ansatz for the structure of the
symmetry which would include symmetries containing the invariant
as a multiplier is most unlikely.  Indeed, {\it a priori} the
likelihood of an Ansatz producing such a multiplier would be so
low that its occurrence would verge on the miraculous! A point
which may be worth future consideration is whether the symmetries
associated with a particular first integral have the
integrals/invariants as multipliers or is there simply no ordained
structure as we proposed for the system  \re{2.2}?  
please consider!) We recall that for the system \re{2.2} we did
not have $I$ but $I^{\ha}$ as the coefficient of $t$, whereas here
we have $I$ and $J$ themselves.

\section{The second three-dimensional system}

\subsection{Painlev\'e analysis}

In \re{4} the variables can be rescaled to give the system
\begin{subequations}
\label{6.1}
\begin{gather}
\dot{u} = vw\n\\
\dot{v} = v (au +v+w)\n\\
\dot{w} = -w (bu +v+w),
\end{gather}
\end{subequations}
where $a = 1-\mu/\nu $ and $b = 1-\nu/\mu $. We assume that $a\neq b $.

In \re{6.1}
we set the variables at
\begin{equation}
u =\alpha\tau^p,\quad v =\beta\tau^q,\quad w =\gamma\tau^r \label{6.17}
\end{equation}
and obtain the
following sets of powers of $\tau $
\begin{equation}
\begin {tabular}{llcc} p- 1 &q+r & &\\
q- 1 &q+p & 2q &q+r\\
r- 1 &r+p &q+r & 2r  \end{tabular}. \label{6.18}
\end{equation}
Evidently one set of dominant behaviours given by $p =q =r = - 1 $.  The
coefficients are determined from the set of equations
\begin{equation}
\begin{split}
-\alpha \z \beta\gamma\n\\
-\beta \z\beta (a\alpha +\beta +\gamma)\n\\
-\gamma \z -\gamma (b\alpha +\beta +\gamma) \label{6.19}
\end{split}
\end{equation}
and we have
\begin{equation}
\alpha = -\f{2}{a -b},\quad \beta +\gamma =  \f{a +b}{a -b},
\quad \beta\gamma = \f{2}{a-b} \label{6.20}
\end{equation}
with particularly unpleasant expressions for $\beta $ and
$\gamma $ which we do not bother to write.

There are two other sets of possible dominant behaviours given by
\begin{equation}
q = -1,\quad r > - 1,\quad p =r\quad\mbox {\rm or}
\quad r = - 1,\quad q > - 1,\quad p =q. \label{6.21}
\end{equation}
There is a symmetry between these two possibilities and
so we just consider one, the second.  Since the powers are required to
be integral, $q $ and $p $ cannot be negative integers as the analysis requires
and so we must make a full substitution.  We take
\begin{equation}
\begin{split}
u\z \sum_{i = 0}a_i\tau^i\n\\
v\z \sum_{i = 0}b_i\tau^i\n\\
w\z \sum_{i = 0}c_i\tau^{i- 1}.\label{6.22}
\end{split}
\end{equation}
On equating coefficients of like powers of $\tau $ to zero we obtain
the following set of information
\begin{equation}
\begin{array}{l|lll}
i & 0 & 1 & 2\\ \hline
a &a_0 &a_1 & (a -b)a_0a_1\\
b & 0 &a_1 &\ha (2a-b)a_0a_1\\
c & 1 & -\ha ba_0&\otw b^2a_0^2-\oth (b+ 1)a_1
\end{array} \label{6.23}
\end{equation}
with the rest
of the coefficients following naturally.  We observe that $a_0 $ and $a_1 $ are
arbitrary and these, in combination with $t_0 $, provide the three arbitrary
constants of integration required to give the general solution of \re{6.1}.
Thus the subselfsimilarity dominant behaviour leads to a general solution and
not a particular (or singular, an unfortunate word in this context) solution.

It remains to look at the resonances of the all terms dominant singular
behaviour.  We set
\begin{equation}
\begin{split}
u \z \alpha\tau^{-1} +\mu\tau^{r- 1}\n\\
v \z \beta\tau^{- 1} +\nu\tau^{r- 1}\n\\
w \z \gamma\tau^{- 1} +\sigma\tau^{r- 1}, \label{6.24}
\end{split}
\end{equation}
observe the following simplifications
\begin{equation}
a\alpha+\beta +\gamma = - 1\qquad b\alpha +\beta +\gamma = 1 \label{6.25}
\end{equation}
and obtain the characteristic equation for the resonances
\begin{equation}
\left|\begin{array}{ccc}
r- 1 & -\gamma & -\beta\\
-a\beta &r-\beta & -\beta\\
b\gamma &\gamma&r+\gamma
\end{array}\right| = 0. \label{6.26}
\end{equation}
We obtain
\begin{equation}
r = -1,2,\beta -\gamma. \label{6.27}
\end{equation}
The first two values for the resonances are
fine, but the third is almost certainly going to be a disaster except for some
very specific values of the parameters $a $ and $b $.  Since the third contains
the difference of the two roots of the quadratic, even for "nice" values of $a
$ and $b $ the value of $r $ could be terrible.

We conclude that
generically \re{6.1} does not possess the Painlev\'e property and is not
integrable in the sense of Painlev\'e.  There is an
amusing contrast with the results of the singularity analysis of the equations
for the Mixmaster universe.  In that analysis the all terms dominant behaviour
was better than the not all terms dominant behaviour
\cite{Contopoulos,Contopoulos2}.

\subsection{First reduction and symmetry analysis}
If we define $r =vw $, in (\ref {6.1}a) and \re{6.2}, we reduce \re{6.1}
to the following two-dimensional system
\begin{equation}
\begin{split}
\dot{u}\z r\n\\
\dot{r}\z (a -b) ru. \label{6.13}
\end{split}
\end{equation}

When we make the Painlev\'e analysis of \re{6.13}, we find that the leading order
behaviour is
\begin{equation}
u = - \f{2}{(a -b)}\tau^{-1} \qquad r = \f{2}{(a -b)}\tau^{- 2}. \label{6.15}
\end{equation}
The resonances occur at $r = - 1, 2 $ and, as one would
expect from a system of such symmetry, there is no incompatibility at the
second resonance.  We obtain
\begin{equation}
\begin{split}
u \z - \f{2}{(a -b)}\tau^{-1} +\ga\tau\n\\
r \z  \f{2}{(a -b)}\tau^{- 2} +\alpha. \label{6.16}
\end{split}
\end{equation}

The Lie point symmetries of \re{6.13} up to the third order in the
variables are
\begin{equation}
\begin{split}
G_1\z \pt\n\\
G_2\z -t\pt+u\pu +r\pr\n\\
G_3\z r\pu + (a-b)ur\pr\n\\
G_4\z t\lb\pt+r\pu + (a -b)ur\pr\rb\n\\
G_5\z u\lb\pt+r\pu + (a-b)ur\pr\rb\n\\
G_6\z r\lb\pt+r\pu + (a -b)ur\pr\rb\n\\
G_7\z 2I\pt\n\\
G_8\z \(2I-(a -b)u^2\)t\pt +u\((a -b)u^2- 4r\)\pu-4r^2\pr. \label{6.14}
\end{split}
\end{equation}
with the simpler system
\re{6.13} it is possible to make the computations at least to
order four in the variables, but there is little point in writing them down as
most of them are of the type of $G_4 $ except with more elaborate coefficients.

\subsection{Symmetries and solution}
The combination (\ref {6.1}b) $w $ $+ $ $v $(\ref {6.1}c)
gives
\begin{equation}
(vw)^.  = (a-b)u (vw) \label{6.2}
\end{equation}
which, when combined with (\ref{6.1}a), becomes
\begin{equation}
\ddot {u} = (a -b)u \dot{u}. \label{6.3}
\end{equation}
We note
that \re{6.3} has only the two symmetries
\begin{equation}
\begin{split}
G_1\z \pt \n\\ G_2\z -t\pt+u\pu. \label{6.4}
\end{split}
\end{equation}

We observe that to integrate the system \re{6.1} it is only necessary to go to
a second order ordinary differential equation.  Equation \re{6.3} is integrated
once to give
\begin{equation}
\dot{u} = I +\ha(a -b)u^2 \label{6.5}
\end{equation}
which immediately
leads to the quadrature
\begin{equation}
\begin{split}
&t-t_0 = \int \f{\d u}{I +\ha(a -b)u^2}\n\\
&\ha (a-b) (t-t_0) = \( \f{a -b}{2I}\)^{\ha}\arctan \(
\f{a -b}{2I}\)^{\ha}u.\label{6.6}
\end{split}
\end{equation}
inversion of \re{6.6} and the trivial quadrature of \re{6.2}
immediately give
\begin{equation}
\begin{split}
&u = A\tan\gW (t-t_0)\n\\
&vw = B\sec^2\gW (t-t_0)\label{6.7}
\end{split}
\end{equation}
in which $A =[2I/(a -b)]^{\ha} $, $\gW =[\ha I (a -b)]^{\ha} $, $B $
is another constant of integration and we have assumed that the parameters
in \re{6.5} are positive.  We consider this case only as the other
possibilities require a similar discussion and this would be repetitive.

We are now left with the integration of (\ref{6.1}c) which now has the simple
form
\begin{equation}
\dot{w} +bA\tan\gW (t-t_0)w+w^2+B\sec^2\gW (t-t_0) = 0. \label{6.8}
\end{equation}
Equation \re{6.8} is a Riccati equation and we transform it to a linear
second order equation by means of the transformation $w = \dot{\eta}/\eta $ to
obtain
\begin{equation}
\ddot {\eta} + bA\tan\gW (t-t_0) \dot{\eta}+B\sec^2\gW (t-t_0)\eta =
0. \label{6.9}
\end{equation}
Equation \re{6.9} can be improved in appearance by the
change of variables
\begin{equation}
\eta (t) =y (x)\qquad x (t) =\tan\gW (t-t_0).\label{6.10}
\end{equation}
We now have
\begin{equation}
\(1+x^2\)y''+\(2 + \f{2b}{a-b}\)xy' + \f{2B}{I(a-b)}y = 0. \label{6.11}
\end{equation}

Equation \re{6.11}, being a linear second order differential equation, has eight
symmetries.  This knowledge is not of much use in the solution of the equation
since seven of
these symmetries require a knowledge of the solution.  The eighth is the
homogeneity symmetry, $y\p/\p y $, which was used to obtain this equation from the
Riccati equation.  Equation \re{6.11} is a little disturbing because it
contains two constants of integration and one would prefer to discuss the
solution with only the parameters, $a $ and $b $, present.  The constants of
integration can be removed by dividing \re{6.11} by $y $ and differentiating
with respect to $x $.  Unfortunately the resulting third order equation looks
too awful to contemplate and that line of action is not pursued here.  In
general the solution of \re{6.11} is in terms of the hypergeometric function
which is a bit too diffuse for our purposes.  There are certain values of the
constants for which solutions are known \cite{Maharaj} for these have been used
in the solution of the problem of the Tikekar superdense stars
\cite{Tikekar,Tikekar2,Tikekar3}.
Kamke \cite[2.261, p 470]{Kamke} provides a method for obtaining polynomial
solutions, again for certain values of the constants.

From the solutions of \re{6.11} we can find $w $ and hence $v $ from \re{6.7}.
Consequently the passing of the Painlev\'e property depends critically on the
solution of \re{6.11}.  On the other hand there is no constraint of this type
on $u $.

The presence of the two parameters $a $ and $b $ in the system \re{6.1} makes
the analysis of the Lie point symmetries of the system problematic.  The
symmetries up to the quadratic symmetries are
\begin{equation}
\begin{split}
G_1\z \pt\n\\
G_2\z -t\pt+u\pu +v\pv +w\pw\n\\
G_3\z vw\pu +v (au +v+w)\pv -w (bu +v+w)\pw\n\\
G_4\z \lb(a -b)u^2- 2vw\rb\pt. \label{6.12}
\end{split}
\end{equation}
In $G_4 $ we recognise that the
coefficient of $\p/\p t $ is simply the first integral in  \re{6.5}.

\section{The third three-dimensional system}

\subsection{Painlev\'e analysis}

We write \re{3} in the equivalent form
\begin{gather}
\dot{u} = (2u+\gl v)w \label{4.1}\\
\dot{v} = (-\gl u+ 2v)w \label{4.2}\\
\dot{w} = u^2+v^2+w^2. \label{4.3}
\end{gather}

We consider the Painlev\'e analysis of the system \re{4.1} -- \re{4.3}.
We find that the powers in the different
terms are
\begin{equation}
\begin{array}{@{}l@{\qquad}l@{\qquad}l@{\qquad}l}
- 1 &r & -p+q+r &\\[0.5ex]
- 1 & -q+p+r &r &\\[0.5ex]
- 1 & 2p-r & 2q-r &r,
\end{array}\label{4.26}
\end{equation}
where we have assumed the leading terms to be
\begin{equation}
\begin{split}
u =\ga\tau^p\n\\
v =\beta\tau^q\n\\
w =\gg\tau^r. \label{4.27}
\end{split}
\end{equation}
The only possible singular behaviour is $p =q =r = - 1 $.  The coefficients
$\ga $, $\beta $ and $\gg $ are determined from the system
\begin{gather}
-\ga = (2\ga +\gl\beta)\gg\label{4.28}\\
-\beta = (-\gl\ga + 2\beta)\gg\label{4.29}\\
-\gg =\ga^2+\beta^2+\gg^2.\label{4.30}
\end{gather}
The combination of $\ga \re{4.28} +\beta \re{4.29} $ gives
\begin{equation}
- \(\ga^2+\beta^2\) = 2\(\ga^2+\beta^2\)\gg\label{4.31}
\end{equation}
from which it follows that either $\gg = -\ha $ or $\ga^2+\beta^2 = 0 $.  We
consider each in turn.

From \re{4.30} it follows that
\begin{equation}
\ga^2+\beta^2 =\oqr.\label{4.32}
\end{equation}
With this value of $\gg $  \re{4.28} and \re{4.29} become
\begin{equation}
\gl\beta = 0\quad \gl\ga = 0\quad\Rightarrow\quad\gl = 0\label{4.33}
\end{equation}
since the assumption of leading order behaviour requires that $\ga $ and
$\beta $ be nonzero.  A similar analysis for the second possibility leads to
the result that $\gl =\pm i $.

Case: $\gl = 0 $
The system is now
\begin{equation}
\begin{split}
\dot{u} \z 2uw\n\\
\dot{v} \z 2vw\n\\
\dot{w} \z u^2+v^2+w^2\label{4.34}
\end{split}
\end{equation}
and to determine the resonances we make the substitution
\begin{equation}
\begin{split}
u \z \ga\tau^{-1} +\mu\tau^{r-1}\n\\
v \z \beta\tau^{-1} +\nu\tau^{r-1}\n\\
w \z \gg\tau^{- 1} +\gs\tau^{r- 1}.\label{4.35}
\end{split}
\end{equation}
The characteristic equation for $r $ is
\begin{equation}
\left|\begin{array}{ccc}
r- 1-2\gg & 0 & - 2\ga\\
0 &r- 1-2\gg & - 2\beta\\
- 2\ga & - 2\beta &r- 1-2\gg
\end{array}\right| = 0\label{4.36}
\end{equation}
which gives the values
\begin{equation}
r =-1, 0, 1\label{4.37}
\end{equation}
when the values of the constants are substituted.  There is only one
eigenvector which is
\begin{equation}
\bfe =\(\begin{array}{r} 2\beta\\- 2\ga\\0\end{array}\).\label{4.38}
\end{equation}
That there is only one eigenvector indicates that it is necessary to
introduce a logarithmic term and the $0 $ for the entry of $w $ indicates
that the logarithmic term must come in with the other two variables, $u $
and $v $.

Case: $\gl =\pm i $
With the same substitution as that in \re{4.35} the characteristic equation
for the resonances is given by
\begin{equation}
\left|\begin{array}{ccc}
r- 1-2\gg & -\gl\gg &  2\ga +\gl\beta\\
\gl\gg &r- 1-2\gg & -\gl\ga + 2\beta\\
- 2\ga & - 2\beta &r- 1-2\gg
\end{array}\right| = 0\label{4.39}
\end{equation}
and with the constraints on the constants gives the values
\begin{equation}
r = - 2, - 1,0.\label{4.40}
\end{equation}
The value $r = 0 $ is the critical one.  Two possible results
occur. In the case that $\ga =\gl\beta $ $\mu $ and $\nu $ are
arbitrary because the rank of the matrix becomes one.  There is no
need to introduce a logarithmic term.  In the case that $\ga \neq
\gl\beta $ this is not the case because $\mu$ and $\nu$ are
related by $2\ga\mu + 2\beta\nu = 0 $.  A logarithmic term must be
introduced.

We conclude that only for the specific values of $\lambda$ shown above does the
system \re{4.1} -- \re{4.3} pass the Painlev\'e test.

\subsection{The two reductions}
We now consider the integration of the system \re{4.1} -- \re{4.3}.
We observe that the
combination $u \re{4.1} +v \re{4.2} $ gives
\begin{equation}
u\dot{u} +v\dot{v} = 2 (u^2+v^2)w \label{4.4}
\end{equation}
so that differentiation of \re{4.3} with respect to time gives the second
order ordinary differential equation
\begin{equation}
\ddot{w} - 6w\dot{w} + 4w^3 = 0 \label{4.5}
\end{equation}
which we recognise as the equation we already met in \re{2.5}.  We know that
it possesses eight Lie point symmetries, that under a Riccati transformation
it becomes the simplest third order equation and that the solution is
\begin{equation}
w = -{}\f{B+Ct}{A+2Bt+Ct^2}.\label{4.6}
\end{equation}

Now that we know the functional form of $w (t) $ we see that \re{4.1} and
\re{4.2} now constitute a linear system.  This system is rendered autonomous
by the introduction of a new time variable \cite{Burgan1, Burgan2} defined by
\begin{equation}
\dm{\tau}{t} = w
\Leftrightarrow \tau = \log\(A+2Bt+Ct^2\)^{-\ha}. \label{4.7}
\end{equation}
If we denote differentiation with respect to $\tau $ by a $'$, the linear
system is
\begin{equation}
\(\begin{array}{c}u\\v\end{array} \)' = \(\begin{array}{rr} 2 &\gl\\-\gl &
2\end{array}\)\(\begin{array}{c}u\\v\end{array} \).\label{4.8a}
\end{equation}
The solution of \re{4.7} is
\begin{equation}
\(\begin{array}{c}u\\v\end{array} \) =\(\begin{array}{c}-i\\1\end{array}
\)\mu\e^{(2+i\gl)\tau} +
\(\begin{array}{c}i\\1\end{array} \)\nu\e^{(2-i\gl)\tau}.\label{4.8}
\end{equation}
There appear to be too many constants of integration, but the three in the
expression for $w $ are in reality two and a small calculation shows that the
constants of integration in \re{4.8}, \viz $\mu $ and $\nu $, are related to
the other constants via
\begin{equation}
2\mu\nu = B^2-AC \label{4.9}
\end{equation}
so that the number of arbitrary constants is in fact three.  In terms of the
variable $\tau $ the solutions for $u $ and $v $ are analytic.  Given the
definition of $\tau $ in \re{4.7}, it is evident that the solution in terms
of the original variable, $t $, is not analytic in general. In terms of the
variable $t $ the solution for $w $ has simple poles as the only singularities.
If we express this solution in terms of the new time, we find that
\begin{equation}
w = \pm\e^{2\tau}\sqrt{B^2-AC+C\e^{-2\tau}} \label{4.10}
\end{equation}
which has a branch point singularity.  Of course, to define a new time in terms
of the solution of one of the variables means that the problem has been reduced
from a three-dimensional system to a two-dimensional system and a linear one at
that.

In fact we can interpose an intermediate system between \re{3} and \re{4.5} by
introducing a new variable, $r $, defined by
\begin{equation}
r^2 = u^2+v^2.\label{4.11}
\end{equation}
The system is
\begin{equation}
\begin{split}
&\dot{r} = 2rw \\
&\dot{w} = r^2+w^2.\label{4.12}
\end{split}
\end{equation}
The solution to the system \re{4.12} is \re{4.6} and
\begin{equation}
r = \f{B^2-AC}{\(A+2Bt+Ct^2\)^2}  \label{4.13}
\end{equation}
in which the additional constant of integration has already been identified in
terms of the three constants found in the solution for the function $w $.

We now consider the Painlev\'e analysis of \re{4.5}, \re{4.12} and \re{3} (in the
present variables) in turn.  In the case of \re{4.5} the only possible singular
behaviour is given by $p = - 1 $, but there are two possible families of
solutions since the coefficient of the leading term can be either $- 1 $ or
$-\ha $.  To determine the resonances we write
\begin{equation}
w = \ga \tau^{-1} +\beta\tau^{r-1}\label{4.17}
\end{equation}
so that we can treat both cases simultaneously.  We find that
\begin{equation}
r = - 1, 6\ga + 4\label{4.18}
\end{equation}
so that we have
\begin{equation}
\ga = -\ha\quad r = - 1,1\qquad \ga = - 1\quad r = - 1, - 2. \label{4.19}
\end{equation}
Evidently the first gives the right Painlev\'e series, that is the Laurent expansion
in the neighbourhood of the singularity,
and the second the left Painlev\'e series, the
asymptotic expansion, that is the Laurent expansion away from the singularity.

For the system \re{4.12} the solution for $w $ is given in \re{4.6} and it is
a simple quadrature to find that
\begin{equation}
r =\f{B^2-AC}{A+2Bt+Ct^2}.\label{4.20}
\end{equation}
We determine the leading order behaviour by writing
\begin{equation}
r =\ga\tau^p\qquad w =\beta\tau^q\label{4.21}
\end{equation}
and substituting this into \re{4.12}.  We have
\begin{equation}
\begin{split}
\ga p\tau^{p-1} \z 2\ga\beta\tau{p+q}\n\\
\beta q\tau^{q- 1} \z \ga^2\tau{2^p} +\beta^2\tau^{2q}.\label{4.22}
\end{split}
\end{equation}
Evidently there are two possible behaviours.  We can have $p =q = - 1 $ or
$p > - 1,q = - 1 $. The roles of $p $ and $q $ cannot be reversed in the
second possibility.  We find that $\ga =\pm\ha $ and $\beta = -\ha $. The
resonances are given by $r =\pm 1 $.  The first two terms of the right
Painlev\'e series are
\begin{equation}
\(\begin{array}{c}
r \\w
\end{array}\) =
\(\begin{array}{c}
\pm\ha \\\ha
\end{array}\)\tau^{-1} +
\(\begin{array}{c}
1 \\\pm 1
\end{array}\)\mu,\label{4.23}
\end{equation}
where $\mu $ is the second arbitrary constant required for the general
solution.  For the second possibility we set
\begin{equation}
r =\sum_{i=0}a_i\tau^i\qquad w =\sum_{i=0}b_i\tau^{i-1}\label{4.24}
\end{equation}
since the standard Painlev\'e analysis breaks down in the case that the
leading order behaviour is not singular in all variables
\cite{Contopoulos,Contopoulos2}.
We substitute \re{4.24} into \re{4.12} to obtain
\begin{equation}
\begin{split}
&\sum_{i=0} i a_i\tau^{i-1} = \sum_{i = 0} \sum_{j = 0} a_i b_j
\tau^{i+j- 1}\n\\
&\sum_{i=0} (i- 1)b_i\tau^{i-2} = \sum_{i = 0}\sum_{j = 0} a_i a_j\tau^{i+j} +
\sum_{i = 0}\sum_{j = 0} b_i b_j\tau^{i+j- 2}.\label{4.25}
\end{split}
\end{equation}
When we compare coefficients of like powers of $\tau $, we find that all
coefficients except $b_0 $ vanish so that the solution is a singular
(particular)
solution and is unrelated to the general solution.  The same result is
found if we commence the series for $r $ at a higher power of $\tau $.
Thus we see that the Painlev\'e property holds equally well for \re{4.12}
as it did for \re{4.5}.  Bearing in mind the results of the Painlev\'e analysis
of the system \re{3} we see that the decomposition of the system
\re{4.12} into the system \re{3} and even the possibility of the preservation
of the Painlev\'e property is restricted to those special cases for the value of
$\gl $ obtained above.

In the case of this system going from a system of two equations to a system
of three equations has had a drastic effect upon the possibility of the
possession of the Painlev\'e property.  Generically, \ie for arbitrary values
of $\lambda$, this system of three
equations does not have the Painlev\'e property.

\subsection{Symmetries}

The Lie point symmetries for \re{4.5} are
\begin{gather}
G_1 = \pt\notag\\
G_2 = -t\pt+w\pw\notag\\
G_3 = w\(\pt+w^2\pw\)\notag\\
\begin{split}
&G_4 = tw\pt+w^2(1+tw)\pw\\
&G_5 = t^2w\pt+w\(1+2tw+2t^2w^2\)\pw
\end{split}\\
G_6 = t^2\(3+2tw\)\pt+\(6t^2w^2+4t^3w^3\)\pw\notag\\
G_7 = t^2\(1+tw\)\pt+tw\(1+3tw+2t^2w^2\)\pw\notag\\
G_8 = t^3\(1+tw\)\pt + t\(1+3tw+4t^2w^2+2t^3w^3\)\pw,\notag \label{4.14}
\end{gather}
those of \re{4.12} up to cubic in the variables are
\begin{equation}
\begin{split}
G_1 \z \pt\n\\
G_2 \z -t\pt+r\pr+w\pw\n\\
G_3 \z 4rw\pr +\(r^2+w^2\)\pw\n\\
G_4 \z t\lb\pt+4rw\pr +\(r^2+w^2\)\pw\rb\n\\
G_5 \z r\lb\pt+4rw\pr +\(r^2+w^2\)\pw\rb\n\\
G_6 \z w\lb\pt+4rw\pr +\(r^2+w^2\)\pw\rb \label{4.15}
\end{split}
\end{equation}
and of the original system, \re{3}, (in the present variables) are
\begin{equation}
\begin{split}
G_1 \z \pt\n\\
G_2 \z -t\pt+u\pu+v\pv+w\pw\n\\
G_3 \z (2u+\gl v)w\pu+(-\gl u+2v)w\pv+(u^2+v^2+w^2)\pw\n\\
G_4 \z t\lb \pt+(2u+\gl v)w\pu+(-\gl u+2v)w\pv+(u^2+v^2+w^2)\pw\rb\n\\
G_5 \z u\lb \pt+(2u+\gl v)w\pu+(-\gl u+2v)w\pv+(u^2+v^2+w^2)\pw\rb\n\\
G_6 \z v\lb \pt+(2u+\gl v)w\pu+(-\gl u+2v)w\pv+(u^2+v^2+w^2)\pw\rb\n\\
G_7 \z w\lb \pt+(2u+\gl v)w\pu+(-\gl u+2v)w\pv+(u^2+v^2+w^2)\pw\rb.\n\\
\label{4.16}
\end{split}
\end{equation}
The symmetries in \re{4.15} and in \re{4.16} are essentially the same with the
latter having an additional symmetry due to the presence of the additional variable.

\section{The four-dimensional system}
\subsection{Painlev\'e analysis}
The parameter $\beta$ in \re{1} is an essential parameter which
cannot be removed
by rescaling. In fact any attempt to remove it will simply lead to the
appearance of another parameter in the other terms. We rewrite the system
as
\begin{subequations}\label{7.1}
\begin{gather}
\dot{u} = u(u-w) - vx \n\\
\dot{v} = v[bu + (b- 2)w] \n\\
\dot{w} =  -w(u - w) - vx \n\\
\dot{x} =  -x[(b - 1)u + (b - 3)w].
\end{gather}
\end{subequations}

We perform the Painlev\'e analysis of \re{7.1}.
In the case of all terms being dominant
the singularity is a simple pole and the coefficients of the leading order
terms
must satisfy the following set of equations
\begin{equation}
\begin{split}
-\ga \z  \ga^2 - \ga\gg - \beta\gg \n\\
-\beta \z  b\ga\beta + (b - 2)\beta\gg \n\\
-\gg \z   -\ga\gg +\gg^2 - \beta\gd \n\\
-\gd \z  -(b - 1)\ga\gd - (b - 3)\gg\gd. \label{7.35}
\end{split}
\end{equation}
Since the leading order behaviour assumes that the coefficients are nonzero,
the second and fourth of \re{7.35} give
\begin{equation}
\begin{split}
&b\ga+(b-2)\gg = -1 \\
&(b - 1)\ga + (b - 3)\gg = 1 \label{7.36}
\end{split}
\end{equation}
from which it follows that either $\ga = \gg = -1$ or
\begin{equation}
\ga + \gg = -2 \quad \mbox{and} \quad \ga + \gg = -1 \label{7.37}
\end{equation}
which is a contradiction. Substitution for the value of $\ga$ and $\gg$
into \re{7.35}
requires that the product $\beta\gg = -1$ and, more importantly, $b = -\tha$.
Thus
the Painlev\'e analysis of \re{7.1} cannot even begin without a
restriction of the
value of $b$. We observe in \re{7.14} and \re{7.15} that this value of
$b$ removes the
possible singularity in the trigonometric term, but we are still left with the
square root to give the solution a branch point singularity.

   Nevertheless for the sake of completeness we analyse the situation for this
particular value of $b$. On making the usual substitution for the leading order
behaviour we obtain the following set of exponents
\begin{equation}
\begin{array}{llll}
                       p-1&  2p&      p+r&  q+s\\
                       q-1&  p+q&     q+r&     \\
                      r-1&  p+r&     2r&   q+s \\
                       s-1&  p+s&     r+s&     \\[1.5ex]
                       -1 & p& r&
\end{array}\label{7.38}
\end{equation}
in which the separated bottom line gives the essence of the content of the
second and fourth lines. When all terms are dominant, the singularity is of
order $-1$.  However, there are various options for subdominant behaviour.
These are listed below as
\begin{equation}
\begin{array}{llll|c}
                     p&   q  &    r&    s  &\mbox{\rm Comment}\\
                     -1&   -1 &  -1&   -1 &\mbox{\rm  generic}\\
                     -1 &\geq 0& \geq 0&  \geq 0&\mbox{\rm plus}\\
                     -1 &  -1  & \geq 0&\geq    0 &\mbox{\rm variants}\\
                     -1 &\geq 0& \geq  0&    -1 &p\leftrightarrow r.
\end{array}\label{7.39}
\end{equation}
We note that the leading order behaviour in the subdominant cases must
contain either $p$ or $r$.

   Only in the case of all terms dominant can we apply the standard
algorithm. The coefficients of the leading order
terms are found from the solution of the system
\begin{equation}
\begin{split}
          -\ga \z  \ga^2 - \ga\gg - \beta\gd \n\\
          -2\beta \z  3\ga\beta - \beta\gd \n\\
               -\gg \z \ga\gg + \gg^2 - \beta\gd \n\\
          -2\gd \z   -\ga\gd + 3\gg\gd  \label{7.40}
\end{split}
\end{equation}
and are
\begin{equation}
                   \ga=-1\quad \gg=-1\quad \beta\gd=-1,    \label{7.41}
\end{equation}
\ie one of the coefficients is arbitrary. The resonance behaviour is determined
by the solution of the system of equations
\begin{equation}
\left(\begin{array}{rrrr}
                 r&\gd& -1&\beta\\
               -3\beta&  2r&\beta& 0\\
               -1&\gg&r&\beta\\
                \gd&0& -3\gd & 2r
\end{array}\)
\(\begin{array}{l}\mu\\\nu\\\gs\\\rho
\end{array}\) =
\(\begin{array}{c}0\\0\\0\\0
\end{array}\).\label{7.42}
\end{equation}
The resonances are given by $r = -1(2), 0, 2$.  The double resonance at $-1$
means that we cannot have the general solution since there will be at most
only three arbitrary constants. For the resonances at $r = 0$ and $r = 2$
 we find that the corresponding vectors are respectively
\begin{equation}
\bfa_0 = \(\begin{array}{r} -1\\\pm 1\\
1\\ \mbox{$\begin{array}{c}-\\+\end{array}$}1\end{array}\)
\quad\mbox{\rm and}\quad
\bfa_2 = \(\begin{array}{r} 1\\\pm\ha\\1\\
\mbox{$\begin{array}{c}-\\+\end{array}$}\ha\end{array}\)\mu. \label{7.43}
\end{equation}
Thus we see that the resonance $r = 0$ indicates the existence of two possible
solutions.  Each of these solutions will contain an arbitrary constant
introduced at the $r = 2$ resonance.  

  In the cases of not all terms being dominant the standard algorithm
cannot be applied and so we must substitute Anz\"atse appropriate to the
particular pattern desired. For the second set of indices in \re{7.39} we write
\begin{equation}
\begin{split}
u \z \sum_{i=0}a_i\tau^{i-1}\n\\
v \z \sum_{i=0}b_i\tau^i\n\\
                       w \z \sum_{i=o}c_i\tau^i \n\\
                       x \z \sum_{i=0}d_i\tau^i \label{6.44}
\end{split}
\end{equation}
and after equating the coefficients of like powers of $\tau$ to zero we
obtain the following set of coefficients
\begin{equation}
\begin{array}{llll}
               a_0=-1&      b_0=0 & c_0=0&       d_0=0 \\
               a_1 = 0   &  b_1 = 0&  c_1 = c_1 &  d_1=0\\
               a_2=\oth c_0 &   b_2=0 & c_2=0 &       d_2=0\\
               a_3=0 &      b_3=0 & c_3=\oth c_1&      d_3=0.
\end{array}\label{7.45}
\end{equation}
Evidently the functions $v$ and $x$ are identically zero for this pattern of
singularity.

For the third and fourth sets of indices in \re{7.39} we obtain the same
expansion as in \re{7.45}. All the solutions for the different possible singularity
structures are particular solutions and, as we expected above, the system
\re{7.1} does not satisfy the requirements of the Painlev\'e test for any of the
values of the parameter $b$.

\subsection{The general solutions}
From the combinations of (\ref{7.1}a) $-$ (\ref{7.1}c), (\ref{7.1}b)$x +
v$(\ref{7.1}d) and (\ref{7.1}a)$w + u$(\ref{7.1}c) we obtain respectively
\begin{gather}
\dot{u} - \dot{w} = u^2 - w^2 \label{7.2}\\
(vx)^. =  (u + w)vx \label{7.3}\\
(uw)^. = -(u+w)vx.\label{7.4}
\end{gather}
From the combination of \re{7.3} and \re{7.4} we obtain the first integral
\begin{equation}
- I^2 = uw + vx \label{7.5}
\end{equation}
and of \re{7.2} and \re{7.4} the first integral
\begin{equation}
J = (u - w)vx, \label{7.6}
\end{equation}
where $I$ and $J$ are constants of integration. In \re{7.5} we have chosen
the label
for the first integral as $-I^2$ for later convenience. The other cases
which we
do not treat here can be treated in a fashion similar to what we are about
to do.

When we substitute \re{7.5} into (\ref{7.1}a), we obtain
\begin{equation}
\dot{u} = u^2 + I^2 \label{7.7}
\end{equation}
which can be integrated by an elementary quadrature to give
\begin{equation}
u(t) = I\tan(t - t_0), \label{7.8}
\end{equation}
where $t_0$ is the constant of integration. By the elimination of $vx$
 between \re{7.5}
and \re{7.6} we obtain a quadratic equation for $w(t)$ which,
when we substitute for
$u(t)$ using \re{7.8}, gives
\begin{equation}
w(t) = \f{1}{\sin 2(t-t_0)}\left\{-I \cos 2(t - t_0) \pm\sqrt{\lb I^2 +
J\sin 2(t - t_0)(1 + \cos(t - t_0))/I\rb}\right\} \label{7.9}
\end{equation}
so that \re{7.6} gives
\begin{equation}
vx = -\ha I\sec^2(t - t_0)\left\{I  \pm\sqrt{\lb I^2 +
J\sin 2(t - t_0)(I + \cos(t - t_0))/I\rb}\right\}. \label{7.10}
\end{equation}

We observe that the parameter, $b$, of the system \re{7.1} does not appear in
the solutions obtained above. The presence of the square root suggests that
contrary to the comment of Golubchik and Sokolov \cite{Golubchik} that the
system \re{7.1}
probably does not pass the Painlev\'e-Kowalevskaya test for generic values of
the parameter is something of an understatement. To determine the functions
$v(t)$ and $x(t)$ we can now treat (\ref{7.1}b) and (\ref{7.1}d) as linear
equations for the two variables. We obtain
\begin{equation}
v(t) =  K\exp\lb\int (bu + (b - 2)w))\d t\rb \label{7.11}
\end{equation}
and
\begin{equation}
x(t) =  \bar{K} \exp\lb - \int((b - 1)u + (b - 3)w))\d t\rb \label{7.12}
\end{equation}
from which it is evident that
\begin{equation}
vx(t) = K\bar{K}\exp\lb\int(u + w)\d t\rb, \label{7.13}
\end{equation}
where the notation $vx(t)$ is meant to indicate $vx$ as a function of time,
which
gives us a neat way to write down the solution without going to the effort of
evaluating the integral in either \re{7.11} or \re{7.12}. Given the
appearance of the
square root in $w$ one would not expect to be able to evaluate either integral
with ease. Using this device we find that
\begin{equation}
v(t) = K\(\f{vx(t)}{K\bar{K}}\)^{b-2}\sec^{2I}(t-t_0) \label{7.14}
\end{equation}
and
\begin{equation}
x(t) = \bar{K}\(\f{vx(t)}{K\bar{K}}\)^{3-b}\cos^{2I}(t-t_0). \label{7.15}
\end{equation}
The exponents in the expressions in \re{7.14} and \re{7.15} definitely
indicate that
one should not expect the system, \re{7.1}, to satisfy the Painlev\'e
test for generic values of the parameter, $b$.

\subsection{Second reduction}

Another way to reduce the order of the system \re{7.1} is by the following
procedure. We define the variables
\begin{equation}
\begin{split}
&s = u - w\n\\
&y = u+w \n\\
&z = vx \label{7.16}
\end{split}
\end{equation}
which, with the use of the system \re{7.1}, satisfy the three-dimensional
system of first order equations
\begin{equation}
\begin{split}
\dot{s} \z sy \n\\
\dot{y} \z s^2-2z \n\\
\dot{z} \z yz. \label{7.17}
\end{split}
\end{equation}

Consider the system \re{7.17}.  When we make the usual substitution
for the leading order behaviour, we require the following terms to balance:
\begin{equation}
\begin{split}
&\ga p\tau^{p-1} = \ga\beta\tau^{p+q} \n\\
&\beta q\tau^{q- 1} = \ga^2\tau^{2p} - 2\gg\tau^r\n\\
&\gg r\tau{r-1} = \beta\gg\tau^{q+r}. \label{7.28}
\end{split}
\end{equation}
From the first and last of \re{7.28} it is obvious that $q = -1$.
If all terms are to be dominant we also have $p= -1$ and $r = -2$. However,
when we solve for the coefficients, we find that the first of \re{7.28}
gives $\beta = -1$ and the third $\beta = -2$ unless both $\ga$ and $\gg$
are zero. This immediately implies that $\beta = 0$ which contradicts the
idea of leading order behaviour.  Consequently we must conclude that the case
of all terms dominant is impossible.


There are two possible sets of subdominant
behaviour. These are $p = -1, r > -2$ and $p > -1, r = -2$. Effectively these
two cases are $p = -1, r = -1$ and $p = 0, r = -2$.

We commence with the second and, since one of the putative solutions is
not singular, we make the ansatz
\begin{equation}
\begin{split}
s \z \sum_{i=0}a_i\tau^{i-1} \n\\
y \z \sum_{i=0}b_i\tau^{i-1} \n\\
z \z \sum_{i=0}c_i\tau^{i-1} \label{7.29}
\end{split}
\end{equation}
and substitute these into \re{7.17}. After a modicum of calculation we find that
\begin{equation}
\begin{array}{lll}
a_0=\pm 1&   b_0=-1 &     c_0=b_1\\[0.5ex]
a_1 = \pm b_1& b_1 = b_1&   c_1 = \mbox{$\frac{4}{5}$}b_1^2\\[0.5ex]
a_2 = \pm\mbox{$\frac{2}{5}$} b_1^2&   b_2= - \mbox{$\frac{1}{5}$} b_1^2&
c_2 = -\mbox{$\frac{3}{5}$}b_1^3\\[0.5ex]
a_3 = \pm\mbox{$\frac{3}{5}$} b_1^3&   b_3 = \mbox{$\frac{8}{5}$} b_1^3&
c_3 = . . .
  \end{array}\label{7.30}
\end{equation}
in which it is quite obvious that there are only two arbitrary constants, the
location of the singularity and the coefficient $b_1$. Consequently the Laurent
series does not represent the full solution in the neighbourhood of the
singularity, but a particular solution of the system \re{7.17}.

For the first possibility of subdominant behaviour we substitute
\begin{equation}
\begin{split}
s \z \sum_{i=0}a_i\tau^i \n\\
y \z \sum_{i=0}b_i\tau^{i-1}\n\\
z \z \sum_{i=0}c_i\tau^{i-2} \label{7.31}
\end{split}
\end{equation}
and find
\begin{equation}
\begin{array}{lll}
a_0 = 0 & b_0 = -2 & c_0 = -1\n\\[0.5ex]
a_1 = 0 & b_1 = 0  & c_1 = 0 \n\\[0.5ex]
a_2 = 0 & b_2 = -2c_2 & c_2 = c_2  \label{7.32}\\[0.5ex]
a_3 = 0 & b_3 =0  & c_3 = 0\n\\[0.5ex]
a_4 = 0 & b_4 =\mbox{$\frac{2}{5}$} c_2^2 & c_4=-\mbox{$\frac{3}{5}$}\n
\end{array}
\end{equation}
with only the even coefficients $b_{2i}$ and $c_{2i}$ being nonzero.
The solution depends
upon two arbitrary constants, but is definitely very particular for the system
since one of the functions is identically zero. It is probably incorrect
to regard
this solution as the general solution for the two functions $y(t)$ and $z(t)$.

Somewhat dismayed by the result displayed in \re{7.31} we are smitten by
a stroke of genius and think to apply a left Painlev\'e series to see
if we can
obtain a solution which does contain the function $s(t)$. Thus we write
\begin{equation}
\begin{split}
s \z \sum_{i=0}a_i\tau^{-i-2} \n\\
y \z \sum_{i=0}b_i\tau^{-i-1}\n\\
z \z \sum_{i=0}c_i\tau^{-i-2} \label{7.33}
\end{split}
\end{equation}
and find
\begin{equation}
\begin{array}{lll}
a_0 = \pm(2b_2 - b_1^2)^{\ha}&b_0 = -2&c_0 = -1\n\\[0.5ex]
a_1 = \pm(2b_2 - b_1^2)^{\ha}b_1&b_1 = b_1&c_1 = b_1\n\\[0.5ex]
a_2 = \pm\ha(2b_2 - b_1^2)^{\ha}(b_1^2 - b_2^2)&b_2 = b_2&c_2 =
\ha\(b_2-b^2_1\)\n\\[0.5ex]
a_3 = \pm\osi(2b_2 - b_1^2)^{\ha}(\mbox{$\frac{24}{5}$}b_1b_2 -2b_1^3)&
b_3 = \mbox{$\frac{9}{10}$}b_1b_2-\ha b_1^3&c_3 = -\mbox{$\frac{1}{5}$}b_1b_2
\label{7.34}
\end{array}
\end{equation}
in which both $b_1$ and $b_2$ are arbitrary. However, this is a Laurent expansion
at infinity and so the expansion is in negative powers of the variable $t$ and
so we have a particular asymptotic solution and not the general solution
of \re{7.17}. We recall that the single third-order equation \re{7.19}
had partial
expansions for both left and right series for one of the values of the
coefficient
of the leading order power. Nevertheless we are inclined to think that this
is possibly the first time that a left Painlev\'e series indicating a
particular
solution has been demonstrated in the case of subdominant behaviour.

By differentiating the second of \re{7.17} and utilising the other two equations
we obtain
\begin{equation}
\ddot{y} - y\dot{y} = s^2y. \label{7.18}
\end{equation}
With a repetition of the same procedure on \re{7.18} we obtain a third order
equation for the variable $y$, \viz
\begin{equation}
y\dddot{y}- \dot{y}\ddot{y} - 3y^2\ddot{y} + 2y^3\dot{y} = 0. \label{7.19}
\end{equation}
Equation \re{7.19} possesses just the two symmetries due to invariance under
time translation and rescaling, even if the computation is carried out for
contact symmetries. If we perform the Painlev\'e analysis on \re{7.19}, we find
that the only admissible singular behaviour is a simple pole. There are two
values for the coefficient of this term, \viz $\ga = -1, -2$.
For the first of these
the resonances are given by $r = -1, 1, 2$ and so, in conjunction with the two
symmetries, \re{7.19} passes the Painlev\'e test.
In the case of the second value
of $\ga$ the resonances are $r = -1, \pm 2$. Consequently there is no possibility of
the Painlev\'e test being satisfied for $\ga = -2$.  If we take the value
$+2$, we
obtain a right Painlev\'e series containing only two arbitrary constants. If
we take the value $-2$, we obtain a left Painlev\'e series also containing only
two arbitrary constants. One of these would be a particular solution in the
vicinity of the singularity and the other a particular asymptotic solution,
both in the sense defined by Cotsakis and Leach \cite{Cotsakis}).

   This immediately raises the question of the integrability of \re{7.19} since
it does not pass the Painlev\'e test for all possible singularity patterns.  If
we divide \re{7.19} by $y^2$, the equation becomes exact and we obtain the first
integral
\begin{equation}
                       -L^2 = \f{\ddot{y}}{y} - 3\dot{y} + y^2, \label{7.20}
\end{equation}
where we have taken the expression for the left-hand side to be more
convenient for later usage. We can rewrite \re{7.20} as the differential equation
\begin{equation}
               \ddot{y}-3y\dot{y}+y^3+L^2y=0.   \label{7.21}
\end{equation}
We observe that the constant of integration, $L$, can be scaled
out of \re{7.21} without affecting the other terms.  We also
observe that \re{7.21} is simply a variation on the equation we
met in \S\S 2 and 3 (\re{2.5} and \re{4.5}). The presence of the
linear term does not affect the number of symmetries, which
remains at eight and so implies that there exists a point
transformation which linearises the equation, although it does
make the transformation look more complex.

To solve \re{7.21}
we introduce the Riccati transformation
\begin{equation}
   y=-\f{\dot{\eta}}{\eta}     \label{7.22}
\end{equation}
to obtain the third order linear equation of maximal symmetry,
\begin{equation}
   \dddot{\eta} + L^2\dot{\eta} = 0    \label{7.23}
\end{equation}
with solution
\begin{equation}
   \eta(t) = A + B \sin L(t - to)         \label{7.24}
\end{equation}
so that
\begin{equation}
   y(t) =  \f{BL \cos L(t-t_0)}{A+B\sin L(t-t_0)}  \label{7.25}
\end{equation}
which, provided $|A/B|\leq 1$, obviously has only poles as singularities.

Consequently we have a counterexample to the claim  by Tabor
\cite[p 330]{Someone} that
the possession of the Painlev\'e property requires that all possible patterns of
singular behaviour pass the Painlev\'e test. 

We can make a few other observations. We have noted that \re{7.21}
possesses eight Lie point symmetries with or without the presence
of the linear term. As in the case of \re{2.5} and \re{4.5},
\re{7.21} without the linear term possesses both a left Painlev\'e
series and a right Painlev\'e series. This is not the case when
the linear term is present.  Then the only possibility is a right
Painlev\'e series.  This does not affect the integrability of
\re{7.21} in terms of functions with only poles as moveable
singularities.  The existence of the same algebra of Lie point
symmetries means that there exists a point transformation of
\re{7.21} which removes the linear term. That there is a change in
the nature of the singularity behaviour means that this
transformation cannot be homeographic. Naturally it must have only
polelike singularities since both solutions possess
   the Painlev\'e property.


   This concludes the discussion of the fourth order system.

\section{Discussion}
From the analyses of the systems of equations \re{1}, \re{2}, \re{3}, \re{4}
and \re{5} several conclusions can be drawn.

  The development of the systems of equations, which in the original paper
\cite{Golubchik} were obtained as
realisations of some kinds of nonassociative algebras,
considered in this paper may be viewed as a process of division of a single
higher order equation. In the case of the two-dimensional quadratic system,
\re{5}, one can easily imagine it as being derived from a second order equation
by a standard process of reduction to a system of two first order equations. It
is not surprising that there was no loss of the Painlev\'e property
in this splitting. In the case of \re{2} the three-dimensional
first order quadratic system
was derived naturally from a third order nonlinear equation. Although this
equation \re{3.2} had only two contact symmetries, reduction of order brought
it to a linear second order equation with eight symmetries and consequently
\re{3.2} must have seven nonlocal symmetries with nice reduction properties. It
is not surprising that \re{3.1} possesses the Painlev\'e property. In the case of the
equations, \re{3} and \re{4}, there is a distinct change in behaviour since the
parameters which are present in the equations are essential parameters and
cannot be removed by rescaling. The basic equation is
really a nonlinear second order equation and one
could imagine this being split into two first order
equations followed by a further split of an arbitrary
nature of one of the equations into two first order equations
containing the arbitrary parameter. This
situation is repeated in the case of \re{1} at a higher order in that one starts
from a third order nonlinear equation and obtains three first order equations
which have the Painlev\'e property and which do not contain a parameter.
The fourth first order equation is obtained by a parameter-dependent split
of one of these three first order equations and the Painlev\'e property is lost.

   One can envisage a situation in which a prediction of the preservation
of the Painlev\'e property can be made. Given an $n$th order scalar ordinary
differential equation possessing the Painlev\'e property, one could expect to
be able to make a system of $n$ first order ordinary differential equations
by some sort of natural reduction and the system would still possess the
Painlev\'e property. Any attempt to have more than $n$ first order equations by
some sort of further splitting may inevitably lead to a loss of the Painlev\'e
property.  It is interesting that the second order equations possessing the
Painlev\'e property, \viz \re{2.5}, \re{4.5} and \re{6.21},
which can be viewed as the
origins of the systems \re{2}, \re{3} and \re{5}, respectively
are essentially the same equation
and that the third order equation from which the system \re{1} was derived is
closely related to that equation. This does pose some interesting questions
about the basis of the work of Golubchik and Sokolov \cite{Golubchik}
from the point of
view of the algebraic structure of scalar ordinary differential equations of the
second and high orders.

   A third question which has only been suggested here is the question of
the algebraic structure of the symmetries of the first order systems. There
seems to be some structure in the symmetries listed.
We note that this structure appears to be
independent of the possession of the Painlev\'e property
which is not surprising. Algebraic properties do not require an analytic basis.
Thus a system can be integrable in the algebraic sense and not in the sense
of Painlev\'e.  However, because symmetry is at the basis of the process of
reduction, a system which is integrable in the sense of Painlev\'e must also be
integrable in the algebraic sense.

   Finally we mention the interesting observation that a left Painlev\'e series
can be found in subdominant behaviour. Related to that is the whole
question of the Painlev\'e integrability of \re{7.19}. The solution
\re{7.25} is manifestly possessed of only poles as singularities. The
equation has the Painlev\'e property for one possible variety of leading
order behaviour, but not for the other.
This makes one wonder just how much is required for one to conclude that
there exists a solution expressible in terms of a Laurent expansion about a
polelike singularity.

\subsection*{Acknowledgements}

PGLL thanks Professor G P Flessas, Dean of the School of Sciences, and Dr S
Cotsakis, Director of GEODYSYC, for their kind hospitality while this work was
undertaken and the National Research Foundation of South Africa and the
University of Natal for a sabbatical grant.

\strut\vfill

\pagebreak

\label{lastpage}


\begin{thebibliography}{99}
\small

\bibitem{bkp?}
Abraham-Shrauner B, Govinder K.S and Leach P.G.L, Integration of Second
Order Equations Not Possessing Lie Point Symmetries, {\it Phys Lett A},
1995, V.203, 168--174.

\bibitem{Burgan1}
Burgan J.-R., Sur des groupes de transformation en physique
math\'ematique: Application aux fluides de \'espace des phases et \`a la
m\'echanique quantique, PhD thesis, Universit\'e d'Orl\'eans, Orl\'eans,
1978.

\bibitem{Burgan2}
Burgan J.-R., Munier A., Fijkalow E. and Feix M.R.,  Groupes de
redimensionnement et transformations canoniques g\'en\'erlis\'ees en Physique
non-lineaire, Probl\`emes Inverses, Sabatier P C ed, Cahiers
Math\'ematiques, Montpellier, 1982, 67--76.

\bibitem{Cairo1}
Cair\'o L., Feix M.R., Hua D.D., Bouquet S. and Dewisme A., Hamiltonian Method
and Invariant Search for 2D Quadratic Systems, {\it J Phys A: Math Gen}, 1993,
V.26, 4371--4386.

\bibitem{Cairo2}
Cair\'o L., Feix M.R. and G\'eronimi C., On the Integrability of the
Three-Dimensional Lotka-Volterra System, in Advances in Systems, Signals,
Control and Computers, Editor V.B. Baji\'c,
IAAMSAD and the South African Branch of
the Academy of Nonlinear Sciences, Durban, ISBN 0-620-23135-1, Vol II,
1998,  66--71.

\bibitem{Cairo3}
Cair\'o L., Llibre J. and Feix M.R., Darboux Method and Search for Invariants
of the Lotka-Volterra and Complex Quadratic Systems, {\it J Math Phys},
1999, V.40, 2074--2091.

\bibitem{Mohammed1}
Chishlom J.S.R. and Common A.K., A Class of Second Order Differential
Equations and Related First Order Systems, {\it J Phys A:Math Gen}, 1987,
V.20, 5459--5472.

\bibitem{Conte}
Conte R.,  Singularities of Differential Equations and Integrability, in
Introduction to Methods of Complex Analysis and Geometry for Classical
Mechanics and Nonlinear Waves, Editors D.\ Benest and C. Fr\oe schl\'e, Editions
Fronti\`eres, Gif-sur-Yvette, 1994, 49--143.

\bibitem{Contopoulos}
Contopoulos G., Grammaticos B. and Ramani A., Painlev\'e Analysis for the
Mixmaster Model, {\it J Phys A: Math Gen}, 1993, V.26, 5795--5799.

\bibitem{Contopoulos2}
Contopoulos G., Grammaticos B. and Ramani A., The Mixmaster Universe
Revisited, {\it J Phys A: Math Gen}, 1994, V.27, 5357--5362.

\bibitem{Cotsakis}
Cotsakis Spiros and Leach P.G.L., Painlev\'e Analysis of the Mixmaster
Universe, {\it J Phys A: Math Gen}, 1994, V.27, 1625--1631.

\bibitem{Mohammed2}
Erwin V.J., Ames W.F. and Adams E., Wave Phenomena: Modern Theory and
Applications, Editors C. Rogers and J.B. Moodie, North Holland, Amsterdam,
1984.

\bibitem{Feix}
Feix M.R., G\'eronimi C., Cair\'o L., Leach P.G.L., Lemmer R.L. and
Bouquet S.\'E.,
On the Singularity Analysis of Ordinary Differential Equations Invariant
under Time Translation and Rescaling, {\it J Phys A: Math Gen}, 1997, V.30,
7437--7461.

\bibitem{Feix2}
G\'eronimi Claude, Feix Marc R. and Leach Peter G.L., Periodic Solutions,
Limit Cycles and the Generalised Chazy Equation, in Dynamical Systems,
Plasmas and Gravitation, Editors P.G.L. Leach, S.\'E. Bouquet, J.-L. Rouet
and E. Fijkalow, Springer-Verlag, Heildelberg, 1999, 327--335.

\bibitem{Mohammed3}
Golubev V.V., Lectures on Analytical Theory of Differential
Equations, Gostekhizdat, Moscow-Leningrad, 1950.

\bibitem{Golubchik2}
Golubchik I.Z. and Sokolov V.V., On Some Generalisations of the Factorisation
Method, {\it Theoret and Math Phys}, 1997, V.110, 267--276.

\bibitem{Golubchik}
Golubchik I.Z. and Sokolov V.V., Operator Yang-Baxter Equations, Integrable
ODEs and Nonassociative Algebras, {\it J Nonlin Math Phys}, 2000, V.7,
184--197.

\bibitem{head}
Head A.K., LIE, a PC Program for Lie Analysis of Differential Equations,
{\it Comp Phys Commun}, 1993, V.77, 241--248.

\bibitem{Hua}
Hua D.D., Cair\'{o} L., Feix M.R., Govinder K.S. and Leach P.G.L., Connection
Between the Existence of First Integrals and the
Painlev\'{e} Property in Lotka-Volterra and Quadratic Systems, {\it %
Proc Roy Soc}, 1996, V.452, 859--880.

\bibitem{Kamke}
Kamke E., Differentialgleichungen: L\"osungsmethoden und L\"osungen, Band I,
10{$^{th}$} Aufl, B.G. Teubner, Stuttgart, 1983.

\bibitem{Kostant}
Kostant B., The Solution to a Generalised Toda Lattice and Representation
Theory, {\it Adv Math}, 1979, V.34, 195--338.

\bibitem{Krause}
Krause J., On the Complete Symmetry Group of the Classical Kepler System,
{\it J Math Phys}, 1994, V.35, 5734--5748.

\bibitem{Leach}
Leach P.G.L., First Integrals for the Modified Emden Equation
$\ddot{q}+\ga(t)\dot{q}+q^n=0$, {\it J Math Phys}, 1985, V.26. 2510--2514.

\bibitem{Leach2}
Leach P.G.L., Hierarchies of Similarity Symmetries and Singularity
Analysis, in Dynamical Systems, Plasmas and Gravitation, Editors P.G.L. Leach,
S.\'E. Bouquet, J.-L. Rouet and E. Fijkalow, Springer-Verlag, Heidelberg,
1999, 304--312.

\bibitem{paper2}
Leach P.G.L., Cotsakis S. and Flessas G.P., Symmetry, Singularities and
Integrability in Complex Dynamics II:
Rescaling and Time-Translation in Two-Dimensional Systems, 2000,
(to appear in {\it J Math Anal Appln}).

\bibitem{paper3}
Leach P.G.L., Nucci M.C. and Cotsakis S., Symmetry, Singularities and
Integrability in Complex Dynamics III: Complete Symmetry Groups of
Nonintegrable Ordinary
Differential Equations, 2000, (submitted to {\it J Nonlin Math Phys}).

\bibitem{Lemmer}
Lemmer R.L. and Leach P.G.L., The Painlev\'{e} Test, Hidden Symmetries and
the Equation $%
y^{\prime\prime}+ yy^{\prime}+ky^{3} = 0$, {\it J Phys A},
1993, V.26, 5017--5024.

\bibitem{Lie}
Lie Sophus, Differentialgleichungen, Teubner, Leipsig, 1891, (reprinted:
Chelsea, New York, 1967).

\bibitem{Maharaj}
Maharaj S.D. and Leach P.G.L., Exact Solutions for the Tikekar Superdense
Star, {\it J Math Phys}, 1996, V.37, 430--437.

\bibitem{Mohammed4}
Mahomed F.M. and Leach P.G.L., The Linear Symmetries of a Nonlinear
Differential Equation, {\it Qu\ae st Math}, 1985, V.8, 241--274.

\bibitem{Mohammed5}
Mahomed F.M. and Leach P.G.L., Contact Symmetries of Second Order
Differential Equations, {\it J Math Phys}, 1991, V.32, 2051--2055.

\bibitem{Mohammed6}
Moreira I.C., Exact Polynomial Invariants for Newtonian Systems, {\it Had
J}, 1984, V.7, 475--492.

\bibitem{head2}
Sherring J., Head A.K. and Prince G.E., Dimsym and LIE: Symmetry Determining
Packages, {\it Math Comp Model}, 1997, V.25, 153--164.

\bibitem{Someone}
Tabor M., Chaos and Integrability in Nonlinear Dynamics, John Wiley,
New York, 1989.

\bibitem{Tikekar}
Tikekar R.,  Exact Model for a Relativistic Star, {\it J Math Phys},
1990, V.31, 2454--2458.

\bibitem{Tikekar2}
Tikekar Ramesh, A Class of Exact Solutions of Einstein Equations in Higher
Dimensional Space-Times, {\it J Indian Math Soc}, 1995, V.61, 37--45.

\bibitem{Tikekar3}
Vaidya P.C. and Tikekar R., Exact Relativistic Models for a Superdense
Star, {\it J Astrophys Astr}, 1982, V.3, 325--334.
\end{thebibliography}
\end{document}